# Enhancing Soft Skills in Network Management Education: A Study on the Impact of GenAI-based Virtual Assistants

Dimitris Pantazatos
*Network Management & Optimal Design Laboratory (NETMODE)*
*National Technical University of Athens (NTUA)*
Athens, Greece
dpantazatos@netmode.ntua.gr

Mary Grammatikou
*Network Management & Optimal Design Laboratory (NETMODE)*
*National Technical University of Athens (NTUA)*
Athens, Greece
mary@netmode.ntua.gr

Vasilis Maglaris
*Network Management & Optimal Design Laboratory (NETMODE)*
*National Technical University of Athens (NTUA)*
Athens, Greece
maglaris@netmode.ntua.gr

*Abstract*—The rapid evolution of technology in educational settings has opened new avenues for enhancing learning experiences, particularly in specialized fields like network management. This paper explores the novel integration of a GenAI-based virtual assistant in a university-level network management course, focusing on its impact on developing students' soft skills, notably critical thinking and problem-solving abilities. Recognizing the increasing importance of these skills in the digital age, our study aims to assess the empirical effectiveness of this artificial intelligence-driven educational tool in fostering these competencies among students.

## I. Introduction

Integrating Generative AI (GenAI) in education is a growing field that promises to change traditional teaching and learning methodologies. GenAI technologies, such as large language models (LLMs) like ChatGPT, are on the frontline of this transformation, offering personalised, interactive, and adaptive learning experiences. These advancements are applicable in addressing students' diverse learning styles and enhancing critical thinking and problem-solving competencies. This study will demonstrate the application of a ChatGPT-based virtual assistant in a network management course in the School of Electrical and Computer Engineering (ECE) of the National Technical University of Athens (NTUA), aiming to assess its impact on developing these essential soft skills.

The relevance of GenAI in education has been highlighted in various studies. For instance, Yilmaz et al. [1] demonstrated that artificial intelligence (AI)-driven tools could significantly boost student engagement and motivation. Singh et al. [2] explored how these technologies facilitate differentiated learning, catering to individual student needs. Furthermore, with their systematic review, Quyang et al. [3] gave insights into improving learning outcomes within STEM education through AI. Despite these advancements, there needs to be more empirical research regarding the effectiveness of GenAI in developing soft skills, especially in specialised courses. This study seeks to fill this gap by examining how the ChatGPT-based virtual assistant can enhance soft skills in higher education.

The paper examines the impact of Gen-AI in terms of students' soft skills improvement and is organized as follows: Beginning with an introduction (Section I) that sets the stage for the research; Section II focuses on relating work, providing background on AI tools in education and their potential impacts. Section III outlines the methodology for creating the virtual assistant. Section IV presents the effect of the virtual assistant in the learning procedure, including an analysis of changes in learners' self-assessment of soft skills and their attitudes towards AI tools. Section V discusses these findings, highlighting their significance in the educational landscape and reflecting on the practical implications of integrating AI into teaching practices. The paper concludes with Section VI, summarizing the key takeaways, addressing the study's limitations, and suggesting directions for future research on the integration and impact of AI tools like ChatGPT in educational settings.

## II. Related Work

In the evolving educational landscape, particularly in network management, the introduction of Generative AI (GenAI) tools like ChatGPT marks a pivotal shift. According to R. Peres et al., GenAIs have a significant impact on academia, so there is a necessity for ethical AI integration and transparency in the usage of AI-generated content [4]. Trust is also an issue that has to be further examined. Exploratory research into students' trust in GenAI, particularly in computer science from S. Haque et al. [5], highlights the interplay between trust in these tools and students' motivation, confidence, and academic performance. This study underscores the importance of fostering a trust-based relationship with technology for improved learning outcomes.

The literature extensively discusses the benefits of incorporating ChatGPT in higher education. Schönberger's research, for instance, underscores ChatGPT's role in customizing the educational experience and enhancing the efficiency of administrative operations [6]. However, it also brings attention to the potential pitfalls associated with its use, such as the perpetuation of existing biases, the spread of misinformation, and the inherent limitations of the AI tool itself. These issues present a considerable challenge, requiring a nuanced approach to deploying ChatGPT in academic settings. Developing effective strategies to address these concerns is imperative to leverage the advantages of ChatGPT while mitigating its risks fully. Another work in the same direction is that of J. Dempere et al., which identifies AI's capabilities in augmenting teaching, learning, and administrative efficiency [7]. Yet, it also acknowledges the need for a comprehensive framework of regulations and ethical standards. These measures are crucial for mitigating the risks of misinformation and inherent biases that could stem from the AI's application. This study also advocates for a carefully calibrated approach to integrating ChatGPT into academic practices, ensuring that its deployment is both responsible and beneficial to the educational community.

Extending this discussion, the study of A. Strzelecki focuses on the psychological drivers behind ChatGPT's acceptance among students, identifying habitual engagement, performance-related expectations, and the enjoyment derived from using the tool as primary motivators [8] —Meanwhile, the study of A. Barret and A. Pack reveals divergent views between students and educators on the appropriateness and preparedness for GenAI's incorporation into the educational process [9]. It emphasizes the need for more clarity in guidelines and for educational institutions to create a structured approach for GenAI's adoption. These insights suggest a complex landscape where educators and students navigate the potential pitfalls of GenAI, advocating for strategic, informed, and ethically grounded integration within academic frameworks.

These diverse studies collectively emphasize the nuanced and multifaceted impact of GenAI tools in network management education. While promising significant educational benefits, their integration into the curriculum demands careful consideration of various ethical, practical, and pedagogical challenges. This necessitates strategic planning and proactive engagement from educators and policymakers to ensure that GenAI tools like ChatGPT are harnessed effectively and responsibly.

Building on the comprehensive insights from the related work, the following section pivots towards the practical application of these findings. We will explore the development of virtual assistants powered by ChatGPT, demonstrating they can be used to enhance the teaching experience.

## III. Virtual Assistant

The subsequent sections will outline the key elements necessary for constructing an impactful prompt and demonstrate their practical application in shaping the virtual assistant.

### A. Basic Aspects of Prompt Engineering

The rise of AI in digital competencies has drawn considerable interest across diverse fields, such as design, scholarly writing, and education. P. Korzynski et al. [10] recognize AI prompt engineering as an emerging digital skill and offer a theoretical basis for crafting effective AI prompts. They introduce a structured methodology known as the AI PROMPT framework to navigate text-to-text prompt engineering, encompassing seven crucial elements for its application. These elements are:

- Articulate the Instruction
- Indicate the Prompt Elements
- Provide Ending Cues and Context
- Refine Instructions to Avoid Ambiguity
- Offer Feedback and Examples
- Manage Interaction
- Track Token Length and Task Complexity

These elements can be considered Gen-AI Model agnostic, as they can be regarded as general aspects of an effective prompt.

The Prompting techniques for GenAI tools like ChatGPT come in two primary forms: few-shot and zero-shot. Few-shot prompting employs limited examples to prompt the AI, generating responses or evaluating alternatives, requiring little data and offering scalability[11]. On the other hand, zero-shot prompting does not necessitate pre-training the AI on specific tasks [12]. Advancements by Brian Lester et al. in "soft prompts" refine the AI's task-specific performance without modifying underlying algorithms [13]. Yet, non-specialists may need help to create effective prompts due to the subtleties involved with large language models, a gap noted by J.D. Zamfirescu-Pereira et al. [14]. This highlights the growing need for more user-friendly prompt-design tools. Echoing this sentiment, M. Sharples proposes a more interactive, conversational model between students and AI, pushing beyond conventional prompt-response routines [15]. The forthcoming section will detail a methodology for constructing a Learning Scenario assistant incorporating these advancements.

### B. Implementing the Virtual Assistant

Leveraging the insights from earlier discussions on prompt engineering, this section introduces a way to create interactive exchanges with GenAI tools. This strategy goes beyond conventional AI interactions by fostering a dialogue-centric environment where prompts and subsequent responses evolve in a dynamic conversation, mimicking human-like exchanges. The methodology facilitates a reciprocal communication flow, transforming GenAI into an active participant in problem-solving. It ensures that the GenAI system receives user inputs and builds on them through a series of progressive, contextually rich dialogues, thus elevating the quality and relevance of outcomes. The methodology that was used for the creation of the virtual assistant consisted of the following key elements:

- Interactive Prompts: These are instructions for the AI to engage with the user, soliciting critical details necessary for task completion, such as creating educational content. The AI conducts a systematic inquiry to capture all pertinent information.

- Follow-up Prompts: User-generated prompts that build on the AI's responses to extract further details or clarify the given information, emulating the natural progression of a human conversation toward a defined goal.

The above elements can be considered crucial for creating prompts for educational settings. These elements can be regarded as to be aligned with the concepts of the work of E. Mollick in the following way [16]:

- Using AI to produce varied examples aligns with the idea of "Interactive Prompts." It involves guiding AI to provide multiple examples to aid student understanding, akin to soliciting detailed information necessary for completing educational tasks.

- "Follow-up Prompts" can relate to engaging AI in conversation to refine examples or explanations. This involves evaluating the AI's output by probing its relevance, accuracy, and variety, much like follow-up questions that delve deeper into a subject.

Leveraging the critical elements outlined earlier, the interactive prompt was implemented to facilitate a rich, AI-mediated dialogue for helping learners solve exercises by promoting critical thinking and problem-solving skills. The development unfolded methodically, encompassing a series of distinct steps:

*1) Step 1 - Crafting the Interactive Prompt:* The interactive prompt was introduced. This step involved presenting a scenario in which the GenAI tool assumes the role of an experienced academic tutor who can help learners solve exercises by promoting critical thinking and problem-solving skills.

*2) Stage 2 - Structured Follow-up Prompts:* The follow-up prompts are constructed as a series of questions, each deepening into the specific requirements of the problem that has to be solved. These prompts facilitate a conversational exchange where the AI gathers incremental details about the subject and the problem learners need to solve.

*3) Stage 3 - Iterative Development and Evaluation:* The procedure underscores an iterative process, where the AI presents initial outputs and actively seeks the user's feedback to refine the answer. This collaborative process mimics a real-world question-and-answer cycle. The LLM is instructed to encourage users to revise and improve the answer to the question produced by the LLM or request regeneration of the output, demonstrating adaptability in the AI's approach.

The final version of the Virtual Assistant is this:

*Role: Assume that you are an experienced academic tutor who can help learners solve exercises by promoting critical thinking and problem-solving skills*

*Task: Your task is to answer learners' questions in a way that can help learners improve their critical thinking and problem-solving skills.*

*First, introduce yourself and let the learners know you will ask them questions to solve an exercise/problem plan tailored to their needs. Ask question 1 from the list below and wait for the learner to respond in a follow-up prompt. Then, move to question 2. Wait for the learner to respond, etc.*

*Questions:*

*1. Subject: What subject would you like to focus on today? (e.g., Mathematics, Physics, History)*

*2. Problem: Could you describe a specific problem or topic that you find challenging in this subject?*

*Follow these steps to answer the topic and to promote the skills:*

*1. Tailored Assistance: Based on the subject and challenge you've identified, you will now go into a detailed explanation or provide a step-by-step guide tailored to your needs. This could include breaking down complex theories, offering real-world examples, or walking you through problem-solving techniques.*

*2. Interactive Dialogue: When you explain, encourage learners to ask questions or ask for clarification at any point. This is an interactive session, and the learner's input is valuable in shaping the direction of the discussion. You'll need to respond to the learner's questions, offering further explanations or perspectives to ensure the learner understands clearly.*

*3. Applying Concepts: To help the learner grasp the concepts better, you should present hypothetical scenarios, case studies, or practical applications related to the topic.*

*4. Visual Aids and Resources: Where helpful, the assistant may suggest resources for further reading or studying, such as articles, textbooks, or online materials.*

*5. Collaborative Problem-Solving: You will guide the learner if the challenge involves solving a specific problem. This could be a collaborative effort where you're encouraged to propose steps or solutions, and the assistant helps refine them.*

*6. Encouragement and Support: Please provide constructive feedback throughout this engagement. You support the learner's learning journey and build confidence in the subject matter.*

The impact of the ChatGPT-based virtual assistant will be presented in the following section.

IV. IMPACT OF THE VIRTUAL ASSISTANT

*A. Methodology*

A pre and post-intervention study assessed the impact of the ChatGPT-based virtual assistant. Initially, the network management course students were asked to complete a questionnaire evaluating their self-perception of critical thinking and problem-solving skills. The ChatGPT assistant was introduced as a supplementary tool in their coursework. The assistant was used to facilitate various exercises, offering a tool for students to engage in problem-solving scenarios and critical analysis. After interacting with the assistant, after four weeks, a post-evaluation questionnaire was administered to evaluate any changes in the students' self-assessment of their soft skills.

*B. Data Collection and Analysis*

The study involved students enrolled in the corresponding course, which was 18. The age of the participants predominantly fell within the 20-25 age group, which accounted for 14 of the responses. Regarding gender distribution, the survey had a more significant representation of males, with 14 male respondents compared to 4 female respondents. The majority reported some level of familiarity, with nine respondents indicating 'Some Experience', six indicating 'Moderate Experience', and three reporting 'Extensive Experience'. This diversity in familiarity levels with digital learning tools sets a varied baseline for the intervention.

Participants preferred digital and interactive methods when asked about their usual methods of gathering information or learning new things. The most common approach, reported by four respondents, involved a combination of online resources, discussions with peers or mentors, and trial and error. Three participants each preferred online resources combined with trial and error and books combined with online resources and trial and error. Additionally, the survey revealed that eight participants regularly use AI-based tools for learning, seven have used them occasionally, and three have never used such tools. Responses were diverse regarding the skills deemed most important for their future careers. Yet, some skills were frequently mentioned, including Technical Proficiency, Problem-Solving Skills, Communication Skills, Teamwork and Collaboration, and Critical Thinking. Other notable skills included Adaptability and Flexibility, Continuous Learning, Innovation and Creativity, Project Management, and Ethical and Social Responsibility. These findings suggest a keen awareness among participants of the diverse skill sets required in the evolving job market and an openness to integrating AI and digital tools in their learning processes.

The primary tool for data collection was the questionnaire provided before and after using the tool. This questionnaire principally consisted of Likert scale questions, ranging from 1 to 5, designed to assess the impact of the ChatGPT assistant on students' core competencies, specifically critical thinking and problem-solving skills. The comparison of pre- and post-intervention responses was central to identifying any significant changes in the student's self-perception and application of these soft skills.

The effectiveness of the ChatGPT assistant was further evaluated in terms of user engagement, response relevance, and the depth of critical thinking it promoted. In the following section, the impact will be presented.

*C. Results*

The study asked participants to self-evaluate their critical thinking and problem-solving skills in pre- and post-intervention surveys. A significant improvement was observed in the self-assessment of these skills following the intervention. The mean score for critical thinking increased from 3.67 in the pre-intervention survey to 4.39 in the post-intervention survey. Similarly, the mean score for problem-solving skills rose from 3.83 to 4.50. Statistical analysis using paired t-tests confirmed the significance of these improvements, with p-values of 0.0053 for critical thinking and 0.0037 for problem-solving skills (confidence level 95%). These results indicate a statistically significant enhancement in the participants' self-evaluation of critical thinking and problem-solving abilities following the intervention (Fig. 1).

| Critical Thinking and Problem-Solving Comparison | | | | |
|---|---|---|---|---|
| Skill | Mean (Pre-Intervention) | Mean (Post-Intervention) | T-Statistic | P-Value |
| Critical Thinking | 3.67 | 4.39 | -3.20 | 0.0053 |
| Problem-Solving | 3.88 | 4.50 | -3.37 | 0.0037 |

Fig. 1. Comparison of Mean Self-Assessment Scores for Critical Thinking and Problem Solving Skills Pre- and Post-Intervention

In evaluating the post-intervention survey, participants rated the usability, helpfulness, and time-saving effects of the ChatGPT-based virtual assistant, along with the extent of time reduction and its primary usage in their learning process. The usability of the ChatGPT assistant was favorably rated, with a mean score of 4.44, indicating a high level of satisfaction. Specifically, nine responses rated it as a 5 (excellent), 6 rated it as 4, and 3 rated it as 3 (average).

Regarding the helpfulness in solving coursework problems, most found the ChatGPT assistant to be 'Helpful' (11 responses), while four respondents found it 'Very Helpful'. There were also two neutral responses and one indicating it was 'Not Helpful'. The mean helpfulness rating stood at 3.83, reflecting a generally positive perception.

Regarding time-saving effects, a significant number of participants (10 responses) reported that the ChatGPT assistant 'somewhat reduced the time' required for their coursework, with four respondents experiencing 'significant time reduction' and four reporting 'no change in time required'. The mean score for the extent of time reduction was 3.67 on a scale where 5 indicates significant time savings.

Participants also described diverse primary uses of the ChatGPT assistant, ranging from asking theory-related questions, understanding complex parts of the theory, code checking, and learning specific programming commands to searching through documentation and assisting with homework questions. These varied responses highlight the multifaceted role of the ChatGPT assistant in enhancing the learning experience.

Finally, the surveys revealed insightful shifts regarding learners' attitudes towards integrating AI tools into their learning process. Initially, learners displayed a relatively high openness to using AI tools, with a mean rating of 4.28 in the pre-intervention survey. This openness slightly decreased post-intervention, with a mean rating of 4.06, indicating a marginal change in attitude but still reflecting a generally cheerful disposition towards AI in education. Interestingly, when it came to the shift in perspective about AI in education after the experience, 8 out of 18 respondents reported a more positive outlook, signifying an enhanced appreciation or understanding of AI's potential in educational settings. In contrast, 7 participants noted no change in their perspective, while three remained undecided. These findings suggest that while the intervention may have bolstered a positive view of AI among specific learners, it also left others unchanged, highlighting the varied impacts of AI integration in educational contexts.

## V. Discussion

This study highlights the significant role of AI tools like ChatGPT in enhancing educational practices, particularly in developing critical soft skills. The positive shift in self-assessed critical thinking and problem-solving skills post-intervention underscores the efficacy of these tools in supporting cognitive skill development. Learners' high ratings for the usability and helpfulness of the ChatGPT-based assistant reinforce its potential utility in educational settings.

The slight decrease in openness to AI tools post-intervention suggests nuanced learner attitudes, yet the overall positive view indicates a readiness to integrate AI into learning. Some learners' shift towards a more favorable perspective on AI post-intervention highlights the impact of direct interaction with these technologies.

However, the study's limited participant number and need for a control group point to areas for improvement. Future research could employ a larger, more diverse sample and a Randomized Control Trial design to assess the impact of AI tools in education more robustly. Such studies would offer deeper insights and guide the integration of AI in future educational strategies, contributing to a dynamic and skill-focused learning environment.

## VI. Conclusion

This study illustrates the effectiveness of AI tools like ChatGPT in enhancing critical thinking and problem-solving skills in an educational setting. Despite the limited participant number and absence of a control group, the overall positive response to AI tools suggests their potential utility in modern education. Future research should focus on larger, more diverse samples and employ controlled study designs to explore further and substantiate these findings. Ultimately, this research underscores the growing relevance of AI in education, highlighting its potential to significantly contribute to the development of essential skills in the digital era.

## Acknowledgement

This work has been funded by the Erasmus+ Scaffolding Online University Learning: Support Systems (SOULSS) project (2022-1-IT02-KA220-HED-000090206). More information can be found at https://soulss.eu/.